\documentclass[conference]{IEEEtran}
\IEEEoverridecommandlockouts
\usepackage{cite}
\usepackage{amsmath,amssymb,amsfonts}
\usepackage{algorithmic}
\usepackage{graphicx}
\usepackage{textcomp}
\usepackage{xcolor}
\usepackage{setspace}
\usepackage{xcolor}
\usepackage{soul}
\usepackage{url}

\def\BibTeX{{\rm B\kern-.05em{\sc i\kern-.025em b}\kern-.08em
    T\kern-.1667em\lower.7ex\hbox{E}\kern-.125emX}}
\begin{document}

\title{Early Malware Detection and Next-Action Prediction\\

}

\author{\IEEEauthorblockN{1\textsuperscript{st} Zahra Jamadi}
\IEEEauthorblockA{\textit{Electrical and Computer Engineering Department} \\
\textit{Concordia University}\\
Montreal, Canada \\
zahrasadat.jamadi@concordia.ca}
\and
\IEEEauthorblockN{2\textsuperscript{nd} Amir G. Aghdam}
\IEEEauthorblockA{\textit{Electrical and Computer Engineering Department} \\
\textit{Concordia University}\\
Montreal, Canada \\
amir.aghdam@concordia.ca}

}

\maketitle

\begin{abstract}
In this paper, we propose a framework for early-stage malware detection and mitigation by leveraging natural language processing (NLP) techniques and machine learning algorithms. Our primary contribution is presenting an approach for predicting the upcoming actions of malware by treating application programming interface (API) call sequences as natural language inputs and employing text classification methods, specifically a Bi-LSTM neural network, to predict the next API call. This enables proactive threat identification and mitigation, demonstrating the effectiveness of applying NLP principles to API call sequences. The Bi-LSTM model is evaluated using two datasets. 
Additionally, by modeling consecutive API calls as 2-gram and 3-gram strings, we extract new features to be further processed using a Bagging-XGBoost algorithm, effectively predicting malware presence at its early stages. The accuracy of the proposed framework is evaluated by simulations.
\end{abstract}


\section{Introduction}
Malware is a term describing a malicious program that is installed on a platform such as a personal computer, harming the user by damaging the system, stealing information, or hosting the system for blackmail purposes \cite{b1}. The number of reported cyberattacks continues to increase, and new malware is produced by attackers. According to 2021 SonicWall Cyber Threat Report, internet of things (IoT) malware attack volume in the first six months of 2021 increased by 59\% compared to the previous year \cite{b2}. Additionally, the AV-TEST Institute indicates that every day, 450,000 new malicious programs (malware) and potentially unwanted applications (PUA) are registered \cite{b3}.


As new families of malicious threats emerge and variants of malware continue to develop, conventional signature-based methods for detecting malware often fall short in identifying a large number of threats due to their reliance on known patterns \cite{b4}. Fortunately, learning-based methods are able to detect malware more efficiently since they can recognize and learn the patterns of previously unseen cyber-attacks \cite{b4}.
Early malware detection and mitigation is an important task, especially for the types of malware that are costly to recover from \cite{b5}. It can save resources, minimize damage and protect sensitive information. One way to detect the malware at its early stage is to continuously monitor the application programming interface (API) calls made by the malware during its run-time and analyze them dynamically \cite{b6}. After early detection, it is desired to block the attack using a proper prediction strategy before it affects the other parts of the system.

A sequence of API calls can be modeled as a natural language processing (NLP) task because of their similarities, e.g., following a specific syntax and grammar, and the importance of context in understanding the meaning of a request \cite{b7}. Several studies have been conducted to detect malware by leveraging NLP techniques. Sundarkumar et al. \cite{b8} used text-mining and topic-mining techniques to use API calls sequence for malware detection. Li et al. \cite{b8} built a joint representation of API calls to depict software behaviors and then implemented a Bi-LSTM model to learn the relationship between API calls in a sequence and performed malware detection. Liu et al. \cite{b9} employed several deep learning-based methods for malware detection based on API calls extracted from Cuckoo sandbox.

In this study, API call sequences are modeled as a natural language construct, and principles of NLP are utilized for early malware detection and next-step prediction. We first propose a framework for detecting malware at its early stage. For this purpose, sequences of API calls are modeled as 2-gram and 3-gram strings and used as new features. Bagging-XGBoost algorithm \cite{b10} is then used for malware detection and feature importance identification.  In the second part of the work, we predict the malware's upcoming action(s). A bidirectional long-short term memory (Bi-LSTM) neural network which is a common method in text classification tasks is used to predict the next API call(s). The proposed method will help proactively detect and block the attack before it can cause any significant damage.  

The rest of the paper is organized as follows. In Section~\uppercase\expandafter{\romannumeral 2}, we describe the datasets and the learning-based approaches used for malware detection and API call prediction. Section~\uppercase\expandafter{\romannumeral 3 } presents the obtained results and effectiveness of the proposed framework in detecting malware and predicting API calls. Finally, in Section~\uppercase\expandafter{\romannumeral 4 }, the contributions are summarized and potential directions for future research in this area are suggested.

\section{Proposed Methodology}

\begin{figure}[htbp]

    \centering
    \includegraphics[width=0.5\textwidth]{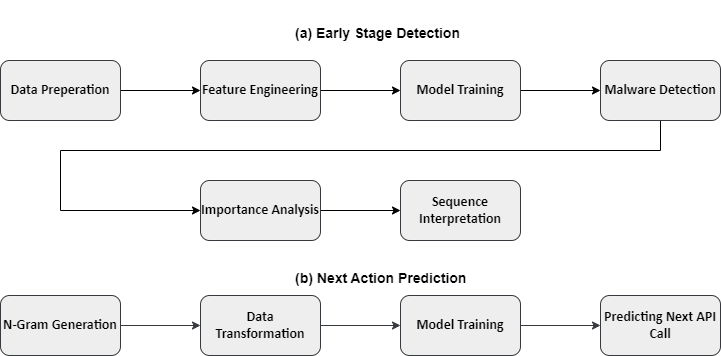}
    \caption{Proposed framework for a) malware detection and b) next-action prediction }
    \label{fig:pipe}
\end{figure}

In this section, we first introduce the two datasets used in this study and will then describe the method used to detect the malware at its early stage and predict its upcoming action(s).
The proposed framework in this study is represented in Figure~\ref{fig:pipe}.

\subsection{Datasets}
The first dataset used in this study contains 42,797 malware and 1,079 goodware API call sequences \cite{b11}. Each sequence contains the first 100 non-repeated consecutive API calls associated with the parent process, with each API call assigned a numerical integer value ranging from 0 to 306. The dataset's large number of malware samples provides a diverse range of malicious behaviors for the models to learn from, while the inclusion of goodware sequences allows the models to differentiate between benign and malicious behavior. This dataset is used for malware detection.

The second dataset comprises 7,107 malware samples from various families, with each malware API call sequence assigned a numerical integer value ranging from 0 to 341 \cite{b12}. This dataset's diversity of malware families enables a comprehensive analysis of the proposed method's effectiveness across different types of malware, while a large number of malware samples ensures that the models are trained on a broad range of malicious behaviors, enabling them to detect and predict various threats.

\subsection{Early Malware Detection}
We use the first dataset \cite{b11} for detecting the malware at its early stage. Due to the importance of early malware detection, API calls for this dataset are only extracted from the parent process, which is primarily responsible for initiating other processes. Furthermore, since the aforementioned dataset is not quantitatively balanced in terms of goodware and malware samples, the random oversampling method is used to increase the number of goodware and balance the dataset for both train and test sets. However, our objective is to also identify malicious activities as the best indicators of the existence of malware. To this end, sequences of API calls, modeled as 2-gram and 3-gram strings of consecutive API calls, are tokenized and used as new features. The most important features are then identified.

Let extreme gradient boosting (XGBoost) algorithm \cite{b13} be used for the binary classification, and subsequently, early detection of the malware. XGBoost is a popular and powerful machine learning algorithm for classification, regression, and ranking tasks. It is an ensemble method that combines the predictions of multiple decision trees to make final predictions \cite{b14}. We then build XGBoost bagging to improve the accuracy and robustness of malware detection tasks. The XGBoost classifier is also used to rank the importance of 2-grams and 3-grams of API calls in terms of their prediction capabilities. Three XGBoost classifiers with learning rates of $ 0.01, 0.05, 0.1$, maximum depth of $4, 3, 5$, and number of estimators equal to $100, 200, 300$, respectively, are used to perform the feature extraction and detection tasks. Hyperparameters for each classifier were selected using grid search.
\subsection{Next Action Prediction}
This subsection presents the main contribution of this work. To the best of the authors' knowledge, no prior research has been reported on predicting upcoming malware actions by predicting the next APIs. We address this problem by modeling the sequence of API calls as a natural language input. We then feed this sequence to Bi-LSTM model to predict the next APIs one by one, which are indicative of the malware's next actions. By predicting the next steps of the attack, proper mitigation techniques can be used to prevent the malware from affecting the other parts of the system. 

Bi-LSTM is a type of recurrent neural network capable of capturing the dependencies between the elements of sequential data. Since it processes the input sequence in both forward and backward directions, it can efficiently identify the words before or after another word \cite{b15}. To ensure that the developed model is able to predict the next API calls of a given sequence efficiently, Bi-LSTM model is tested on both datasets. The N-gram method is applied to both datasets to convert the API calls sequence into a feature structure which helps the Bi-LSTM model learn the relationship of the API calls with each other \cite{b16}. 

To build N-gram features for a sequence of API calls, we consider a set of $n$ consecutive API calls, where \textit{n} ranges from $2$ to the length of the sequence minus $1$. For each subsequence, we use the last API call as the label. For example, to generate the first data point, we take the first two API calls in the sequence and use the third API call as the label. 

An example of N-gram features for a subsequence of one of the malware samples containing $7$ API calls is given in Table~\ref{tab: 1}.

\begin{table}[h]
    \centering
    \caption{N-grams for a subsequence of $7$ numerical API calls.}
    \label{tab: 1}
    \begin{tabular}{|l|l|}
        \hline
        Data point & Corresponding label \\
        \hline
        $[220, 233]$   & $[237]$\\
        $[220, 233, 237]$ & $[220]$ \\
        $[220, 233, 237, 220]$ & $[233]$ \\
        $[220, 233, 237, 220, 233]$ & $[290]$ \\
        $[220, 233, 237, 220, 233, 290]$ & $[260]$ \\

        \hline
    \end{tabular}
    
    \label{tab:mytable}
\end{table}
N-grams can capture the patterns in the sequence of API calls, indicating certain behaviors or outcomes. We can then use these N-gram features to train the Bi-LSTM model for predicting the next API call in a given sequence. The predicted API call is then added to the input sequence and fed to Bi-LSTM network. In this way, we are able to predict multiple API calls which indicates the future action of a malware.

The proposed Bi-LSTM neural network has four layers. The first one is an embedding layer for converting the input sequence into a dense vector capturing its semantic and contextual information. The second one is a Dropout layer with a $30\%$ rate to prevent the model from overfitting. The third one is a Bi-LSTM layer with a size of 150 neurons. Finally, the last one is a dense layer which completes the classification task. Adam optimization algorithm with a learning rate of $0.01$ is employed. The cost function used in this network is categorical cross-entropy, which is widely used in text classification problems. Aforementioned configuration and hyperparameter values are obtained after several experiments to achieve the highest level of performance.




\section{Experimental Results}
\begin{figure}[h]
    \centering
    \includegraphics[width=0.4\textwidth]{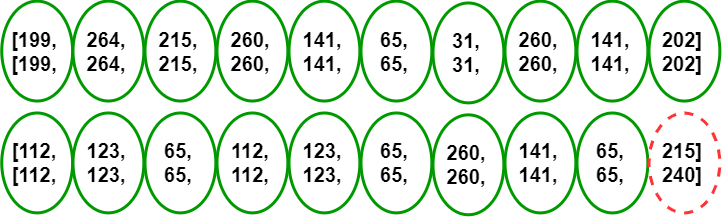}
    \caption{Next $10$ API calls predicted by Bi-LSTM model compared to the ground truth. Input sequences are randomly chosen from the first dataset \cite{b11}. The green circle and red dashed circle indicate the correct prediction and wrong prediction respectively.  }
    \label{fig:2}
\end{figure}

\begin{figure}[h]
    \centering
    \includegraphics[width=0.4\textwidth]{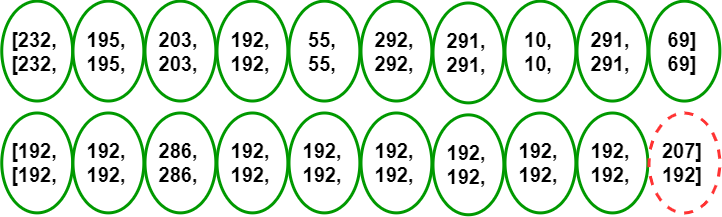}
    \caption{Next $10$ API calls predicted by Bi-LSTM model compared to the ground truth. Input sequences are randomly chosen from the second dataset \cite{b12}. Green circle and red dashed circle indicate the correct prediction and wrong prediction respectively.}
    \label{fig:3}
\end{figure}

The proposed approach is able to predict the upcoming actions of a malware by predicting the next API calls that are going to be made by the malware one at a time. 
Figures \ref{fig:2} and \ref{fig:3} represent the performance of the Bi-LSTM model in predicting the next $10$ API calls for $2$ given input sequences. Input sequences are chosen randomly from the test sets belonging to the first \cite{b11 } and second dataset \cite{b12}. In both figures, the sequence presented on the top is the predicted sequence, while the one presented below represents the ground truth.

\begin{figure}[h]
    \centering
    \includegraphics[width=0.3\textwidth]{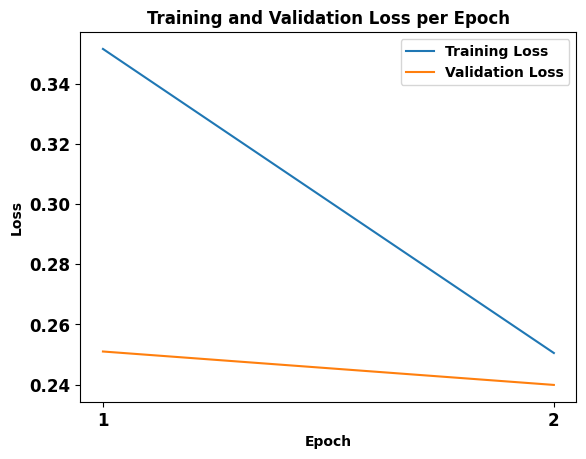}
    \caption{Training and validation loss of Bi-LSTM network on the first dataset \cite{b11}}
    \label{fig:4}
\end{figure}

\begin{figure}[h]
    \centering
    \includegraphics[width=0.3\textwidth]{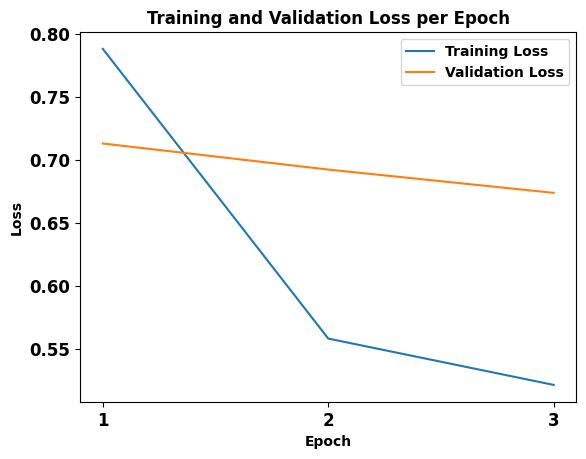}
    \caption{Training and validation loss of Bi-LSTM network on the second dataset \cite{b12}}
    \label{fig:5}
\end{figure}

During the training stage, early stopping method was implemented to determine the optimal number of epochs and prevent overfitting. Figures \ref{fig:4} and \ref{fig:5} show the training and validation loss on each dataset during the training process.

The performance of the Bi-LSTM model for each dataset is evaluated by measuring the accuracy, precision, recall and F1 score which are reported in Table~\ref{tab:2}. In this problem, label imbalance is a concern due to the unequal occurrence of different API calls when a process is running. Some API calls may be more common in both benign and malicious programs, while others might be rare. To overcome this challenge, the weighted average of all labels is reported, which takes the number of true instances for each label into account. Reported results show that the model is performing more accurately in predicting the next API call for the first dataset compared to the second one. This might be due to two factors. First, the first dataset \cite{b11} contains a more diverse repository of malware API call sequences (42797 samples) compared to the second dataset \cite{b12} (7107 samples). This diversity can help the model to generalize more efficiently to unseen patterns of API call sequences. The second reason could be related to the length of API sequence. The first dataset \cite{b11} contains the first 100 non-repeated consecutive API calls and the second dataset \cite{b12} contains all the API calls made during the run-time. The Bi-LSTM model is performing better when trained on shorter sequences of API calls.

\begin{table}[h]
    \centering
    \caption{Accuracy, precision, recall and F1 score of the Bi-LSTM model predicting the next API call using dataset 1 [11] and dataset 2 [12].}
    \label{tab:2}
    \begin{tabular}{|l|l|l|}
         \hline
         & \textbf{Dataset 1} & \textbf{Dataset 2} \\
         \hline
         \textbf{Accuracy} & $93.62\%$ & $88.80\%$ \\
         \hline
         \textbf{Precision} & $93.58\%$ & $88.50\%$ \\
         \hline
         \textbf{Recall} & $93.62\%$ & $88.80\%$ \\
         \hline
         \textbf{F1 score} & $93.52\%$ & $88.48\%$ \\
         \hline
    \end{tabular}
\end{table}




    

In order to evaluate the prediction ability of the Bi-LSTM network, ROC score was calculated for each API call label. ROC score is a common evaluation metric for classifiers. Performance of the classifier can be evaluated by measuring the area under the ROC curve which is a plot of true positive rate (TPR) vs. false positive rate (FPR). By calculating the ROC score for each label, we were able to identify for which API calls the Bi-LSTM network was struggling to make predictions. We then realized for both datasets, these APIs are the ones that are not commonly called during the run-time of the samples and the model struggled to accurately predict these rare API calls when presented with new, unseen data samples. Table \ref{tab:4} presents API functions that are rarely used in two datasets, but are present in both.

\begin{table}[h]
    \centering
    \caption{List of Rare API Calls}
    \label{tab:4}
    \begin{tabular}{|c|l|}
        \hline
        \textbf{No.} & \multicolumn{1}{c|}{\textbf{API Call}} \\
        \hline
        1 & CopyFileW \\
        \hline
        2 & GetUserNameExA \\
        \hline
        3 & GetDiskFreeSpaceExW \\
        \hline
        4 & SHGetSpecialFolderLocation \\
        \hline
        5 & HttpSendRequestA \\
        \hline
        6 & InternetGetConnectedState \\
        \hline
        7 & sendto \\
        \hline
        8 & RtlDecompressBuffer \\
        \hline
    \end{tabular}
\end{table}



For early detection of malware, we generate new features containing two and three consecutive API calls extracted from the parents' process. We then extract 10 most important features identified by the XGBoost classifier and compared their occurrence in malware samples and benign samples. Table~\ref{tab: 5} indicates the most important features and their occurrence in goodware and malware samples. Class $0$ and class $1$ correspond to bengin and malware samples respectively. The features that are more often in malware samples are then investigated to identify any potential malicious activities that occur during the run-time of malware. Table \ref{tab: 6} presents these features ordered by their importance. 
\begin{table}[ht]
\centering
\caption{Top 10 Important sequences of API call sequences and corresponding class frequencies}
\label{tab: 5}
\begin{tabular}{|c|c|c|c|c|}
\hline
\textbf{API Calls} &  \textbf{Class 0 Count} & \textbf{Class 1 Count} \\ \hline
[240, 117, 82] & 1779 & 12287 \\ \hline
[117, 297, 199] & 0 & 6345 \\ \hline
[35, 208, 240] & 625 & 14365 \\ \hline
[199, 264] & 17349 & 16475 \\ \hline
[215, 37] & 549 & 5129 \\ \hline
[114, 215] & 14582 & 3064 \\ \hline
[117, 215, 260] & 12165 & 934 \\ \hline
[215, 37, 158] & 397 & 4993 \\ \hline
[202, 260] & 13248 & 8108 \\ \hline
[117, 215, 89] & 44 & 2276 \\ \hline
\end{tabular}
\end{table}

\begin{table}[h]
\centering
\small
\caption{Top 5 important malicious API call sequences indicating the existence of malware }
\label{tab: 6}
\begin{tabular}{|c|c|}
\hline
\textbf{} & \textbf{ API Call Sequence}  \\ \hline
1& LdrLoadDll, LdrGetProcedureAddress, CertOpenSystemStoreA \\ \hline
2 & LdrGetProcedureAddress, NtCreateFile, SetFilePointer  \\ \hline
3 & GetSystemMetrics, NtAllocateVirtualMemory, LdrLoadDll  \\ \hline
4 & NtClose, NtOpenKeyEx, NtQueryValueKey  \\ \hline
5 & LdrGetProcedureAddress, NtClose, NtDuplicateObject  \\ \hline
\end{tabular}
\end{table}

The malware corresponding to the first API call sequence specified in Table \ref{tab: 6} is to load a malicious Dynamic Link Library (DLL) into memory, retrieve the address of a specific function within the DLL, and obtain a certificate from the system certificate store. The second suspicious API call sequence is when the malware locates a specific function within a loaded module, creates a new file on the system, and manipulates the file pointer to write data to a specific location within the file. The third sequence is extracting system information, allocating memory in the virtual address space, and loading a DLL into memory. The fourth is manipulating the Windows Registry, which is a hierarchical database storing configuration settings and other system information. The last one is locating and duplicating a handle to an object, such as a file, registry key, or process, in order to gain access to it and perform some malicious actions.

Bagging-XGboost is then used to classify an ongoing process as malware or goodware. Table \ref{tab: 7} reports the classification performance of XGBoost model. The results show that the proposed malware detector framework is able to detect the malware at its early stage with high accuracy. The precision score of $92.70\%$ shows that the model has a false alarm rate of $7.3\%$ which is indicative of the well performance of our early malware detector.

\begin{table}[h]
    \centering
    \caption{Accuracy metrics of XGBoost bagging model}
    \label{tab: 7}
    \begin{tabular}{|l|l|}
         \hline
        Accuracy   & $95.85\%$\\
        \hline
        Precision & $92.70\%$\\
        \hline
        Recall & $99.56\%$\\
        \hline
        F1 score & $96.00\%$\\
        \hline

        \hline
    \end{tabular}
    
    \label{tab:mytable}
\end{table}

\section{Conclusions and Future Work}
This paper presents a framework for early-stage malware detection and mitigation by treating API call sequences as natural language inputs and employing text classification methods, specifically a Bi-LSTM neural network, to predict the next API call. This study demonstrates that Bi-LSTM, a neural network architecture commonly used in NLP tasks, is an effective method for predicting API calls due to the similarities between the sequence of API calls and natural language structure [6]. The model is able to predict the next action of the malware by predicting the next API calls that are most likely to occur, one at a time, and allow early mitigation. Additionally, by modeling consecutive API calls as 2-gram and 3-gram strings, we extract new features to be further processed using a Bagging-XGBoost algorithm. This enables us to identify the sequence of API calls and their corresponding activities that may suggest a malware exists.

For future work, one can investigate alternative NLP techniques, such as transformers and attention mechanisms, to enhance the malware detection and prediction capabilities of the framework. Moreover, evaluating the real-time performance of the proposed framework for online malware detection and mitigation could provide insights into its potential for practical deployment in real-world cybersecurity scenarios. Finally, this study only explored the prediction of API calls one step at a time. One direction could involve extending the framework to multistep-ahead prediction of API calls.


\begin{thebibliography}{00}


\bibitem{b1} T. Gržinić and E. B. González, "Methods for automatic malware analysis and classification: a survey," \textit{International Journal of Information and Computer Security}, vol. 17, no. 1-2, pp. 179-203, 2022.
\bibitem{b2} SonicWall, "SonicWall 2021 Cyber Threat Report," 2021. [Online]. Available: \url{https://www.sonicwall.com/resources/2021-sonicwall-cyber-threat-report/}. [Accessed: Apr. 5, 2023].

\bibitem{b3} AV-TEST GmbH, "AV-TEST Statistics," 2023. [Online]. Available: \url{https://www.av-test.org/en/statistics/malware/}. [Accessed: Apr. 5, 2023].

\bibitem{b4} P. Maniriho, A. Mahmood, and M. J. Chowdhury, "Evaluation and survey of state of the art malware detection and classification techniques: Analysis and recommendation," \textit{SSRN Electronic Journal}, 2022.

\bibitem{b5} M. Rhode, P. Burnap, and K. Jones, "Early-stage malware prediction using recurrent neural networks," \textit{Computers \& Security}, vol. 77, pp. 578-594, 2018.

\bibitem{b6} K. Chang, N. Zhao, and L. Kou, "A Survey on Malware Detection based on API Calls," in \textit{2022 9th International Conference on Dependable Systems and Their Applications (DSA)}, Aug. 2022, pp. 464-471.
\bibitem{b7} Nordic APIs, "Should You Design Natural Language First APIs?," \textit{Nordic APIs}, Oct. 18, 2018. [Online]. Available: \url{https://nordicapis.com/should-you-design-natural-language-first-apis/}. [Accessed: Apr. 5, 2023].
\bibitem{b8} G. G. Sundarkumar et al., "Malware detection via API calls, topic models and machine learning," in \textit{2015 IEEE International Conference on Automation Science and Engineering (CASE)}, Aug. 2015, pp. 1212-1217.
\bibitem{b9} C. Li et al., "A novel deep framework for dynamic malware detection based on API sequence intrinsic features," \textit{Computers \& Security}, vol. 116

\bibitem{b10} X. Deng et al., "Bagging-XGBoost algorithm based extreme weather identification and short-term load forecasting model," \textit{Energy Reports}, vol. 8, pp. 8661-8674, 2022.

\bibitem{b11}A. Oliveira and R. Sassi, "Behavioral Malware Detection Using Deep Graph Convolutional Neural Networks," \textit{TechRxiv, preprint}, 2019. [Online]. Available: https://doi.org/10.36227/techrxiv.12020416.v1. [Accessed: Apr. 5, 2023].

\bibitem{b12} F. O. Catak and A. F. Yazı, "A Benchmark API Call Dataset for Windows PE Malware Classification," \textit{arXiv preprint arXiv:1905.01999}, 2019.

\bibitem{b13} T. Chen and C. Guestrin, "Xgboost: A scalable tree boosting system," in \textit{Proc. 22nd ACM SIGKDD Int. Conf. Knowl. Discov. Data Min.}, 2016, pp. 785-794.

\bibitem{b14} C. Bentéjac, A. Csörgő, and G. Martínez-Muñoz, "A Comparative Analysis of Gradient Boosting Algorithms," \textit{Artificial Intelligence Review}, vol. 54, no. 3, pp. 1937-1967, 2020.

\bibitem{b15} K. Zeberga, M. Attique, B. Shah, F. Ali, Y. Z. Jembre, and T.-S. Chung, "A Novel Text Mining Approach for Mental Health Prediction Using Bi-LSTM and Bert Model," \textit{Computational Intelligence and Neuroscience}, vol. 2022, pp. 1-18, 2022.

\bibitem{b16} Y. Zhang and Z. Rao, "n-BiLSTM: BiLSTM with n-gram Features for Text Classification," in \textit{2020 IEEE 5th Information Technology and Mechatronics Engineering Conference (ITOEC)}, Chongqing, China, 2020, pp. 1056-1059, doi: 10.1109/ITOEC49072.2020.9141692.







\end{thebibliography}
\end{document}